\documentclass{article}
\begin{document}

\begin{center}
\centerline{\large \bf  Time invariance violation is a physical base }

\centerline{\large \bf of atomic Bloch oscillations in an optical lattice}
\end{center}

\vspace{3 pt}
\centerline{\sl V.A.Kuz'menko\footnote{Electronic 
address: kuzmenko@triniti.ru}}

\vspace{5 pt}

\centerline{\small \it Troitsk Institute for Innovation and Fusion 
Research,}
\centerline{\small \it Troitsk, Moscow region, 142190, Russian 
Federation.}

\vspace{5 pt}

\begin{abstract}

The physical mechanism of phenomenon is explained as a result of inequality 
of forward and reversed processes in optics. The importance of experimental 
study of its invariance relation is discussed.

\vspace{5 pt}
\end{abstract}

\vspace{12 pt}

There is rather usual situation in nonlinear optics when a 
phenomenon has mathematical description, but clear explanation of its 
physical mechanism is absent. Clear physical sense appears if we use the 
concept of inequality of forward and reversed processes in optics. Such 
approach has quite reliable basis [1]. 

In this note we shall discuss the situation with explanation of the 
Bloch oscillation of ultracold atoms in a vertical optical lattice, 
formed by two counterpropagating beams [2 - 4]. Under action of gravity 
the atoms perform a cyclic motion in the vertical direction. The 
measurement of oscillating frequency allows a sensitive determination of 
the acceleration of gravity or forces at the micrometer length scale. 

According to the existing theoretical concept the physical base of the Bloch 
oscillations is the polarization interaction of the atoms with the standing 
optical wave. The corresponding theoretical description is very complex 
and non-intuitive [5, 6]. It gives for the Bloch oscillations frequency 
following expression:
\begin{equation}
\ \nu_{B} = \frac{mg \lambda}{2h}   
\end{equation}   
where \emph{m} is the atomic mass, \emph{g} is the acceleration of gravity, 
$ \lambda $  is the wavelength of the light, and \emph{h} is 
Planck constant [2]. 

Below we shall deduce the formula (1) by very simple and quite intuitive way.
We also shall give the alternative physical explanation of origin of the 
phenomenon.

When the magneto-optical trap is switched off, the ultracold atoms start 
to fall down under action of the gravity (Fig.1). At the certain moment of 
time the Raman optical transition takes place and the atoms receive a double 
recoil moment in the opposite direction. The obtained recoil energy 
$ (2E_{R}) $ allows for atoms to return in the initial point. The 
amplitude (height) of atomic oscillations (H) may be deduced from condition 
\begin{math}  \ E_{R} = h^{2}/2m\lambda^{2} = mgH \end{math} and it 
is given by 
\begin{equation}
\ H=\frac{h^{2}}{2m^{2}\lambda^{2}g}
\end{equation}
For $^{88}Sr$ and $\lambda$ = 532 nm this height equals $3,66 \mu m$ or 
$6,88 \lambda$ [2]. For $ ^{40}K $ and $\lambda = 873 nm$ it equals 
$6,57 \mu m$ or $7,5 \lambda$ [3,4]. But for $ ^{6}Li $ atoms this 
value should reach $292 \mu m$ or $334,5 \lambda$ (for $\lambda = 873 nm$). 

The fall time (t) of atoms is related to the height by  $H = gt^{2}/2$ . 
So, for the period of oscillations (T) we obtain:
\begin{equation}
\ T=2t=\frac{2h}{mg\lambda}=\frac{1}{\nu_{B}}
\end{equation}        
The equations (1) and (3) are identical.

For the explanation of physical origin of atom oscillations we should 
answer for two questions:

1) Why does the Raman transition occur only at the certain moment of time?

2) Why is the Raman recoil momentum directed only in one side? \\
The answer on these questions is that we deal with a direct 
manifestation of a time reversal noninvariance or inequality of forward 
and reversed processes in optics. This situation is completely the same 
as in the case of splitting and mixing of a photons [7, 8]. Those 
experimental results clearly show that the reversed optical process is 
much more efficient, than the forward one. In our case the Raman transition, 
which return the atom into the initial point is the reversed transition. 
In contrast, the transition, which recoil pushes the atom down, is the 
forward Raman transition. The experimental results show, that the 
cross-section of reversed transition exceeds in many orders of magnitude 
the cross-section of forward transition.

Because of a recoil energy of Raman transition is fixed, the moment of 
time for the reversed transition in the discussed case should be also fixed.

It is clear, that the important task now should be an experimental 
evaluation and measurement of a ratio of cross-sections for forward and 
reversed transitions. Different methods may be used for this goal. 
The most simple and natural way is the using of pump-probe method with 
weak, collinear, copropagating probe laser pulse. The idea of such 
experiments was discussed early [9 - 11]. It is surprising that the 
results of such simple experiments are absent in literature till now.

In conclusion, we propose the alternative physical explanation of origin 
of atomic Bloch oscillations in an vertical optical lattice. This 
explanation is based on the concept of inequality of forward and 
reversed processes in optics. The importance of experimental study of 
the invariance relation of forward and reversed transitions is discussed 
once again.

\end{document}